# Soft Elasticity Optimises Dissipation in 3D-Printed Liquid Crystal Elastomers

D. Mistry[1,2]\*, N. A. Traugutt[1], B. Sanborn[3], R. H. Volpe[4], L. S. Chatham[4], R. Zhou[1], B Song[3], K. Yu[1], K. Long[3], C. M. Yakacki[1,4]\*

[1]Department of Mechanical Engineering, University of Colorado, Denver, Denver, CO, 80204, USA
[2]Present address: School of Physics and Astronomy, University of Leeds, Leeds, LS2 9JT, UK
[2]Materials and Failure Modeling Department, Sandia National Laboratories, Albuquerque, NM, 87123, USA
[4]Impressio Inc., 12635 E. Montview Blvd, Suite 214, Aurora, CO 80045, USA

**Abstract**

Soft-elasticity in monodomain liquid crystal elastomers (LCEs) is promising for impact-absorbing applications where strain energy is ideally absorbed at constant stress. Conventionally, compressive and impact studies on LCEs have not been performed given the notorious difficulty synthesizing sufficiently large monodomain devices. Here, we demonstrate 3D printing bulk (>cm$^3$) monodomain LCE devices using direct ink writing and study their compressive soft-elasticity over 8 decades of strain rate. At quasi-static rates, the monodomain soft-elastic LCE dissipated 45% of strain energy while comparator materials dissipated less than 20%. At strain rates up to 3000 s$^{-1}$, our soft-elastic monodomain LCE consistently performed closest to an ideal-impact absorber. Drop testing reveals soft-elasticity as a likely mechanism for effectively reducing the severity of impacts – with soft elastic LCEs offering a Gadd Severity Index 40% lower than a comparable isotropic elastomer. Lastly, we demonstrate tailoring deformation and buckling behavior in monodomain LCEs *via* the printed director orientation.

**Introduction**

One of the most exciting but often overlooked applications for liquid crystal elastomers (LCEs) is for use in strain-rate-dependent impact absorbing devices.[1–5] In 2001, Clarke *et al*. reported that LCEs – which incorporate the anisotropic ordering of liquid crystals into elastic polymer networks – demonstrate elevated loss tangents ($tan(\delta) = \text{G}''/\text{G}$) as high as 1.5 when held at temperatures between their glass transition and nematic-to-isotropic transition temperatures ($T_\text{g}$ and $T_\text{NI}$ respectively).[6] These values of $tan(\delta)$ correspond to highly viscous and dissipative materials and are far greater than values (~0.1) found in traditional isotropic elastomers.[7,8]

Despite this exceptional mechanical behavior, there have only been a handful of published papers concerning LCE dissipative mechanical properties.[1,2,4,6,9–11] Azoug *et al.* reported on the large and strain-rate dependent tensile hysteresis between loading and unloading stress-strain curves for polydomain (macroscopically unaligned) LCEs.[9] This hysteretic behavior describes an efficient dissipator of strain energy in a material that can either be plastic or elastic depending on the network structure.[5] Moreover, using digital light processing 3D printing, our group recently showed that polydomain LCEs had superior energy dissipation and rate dependency behavior compared to conventional isotropic elastomeric materials in compression.[1] Curiously, this work showed that the well-known soft-elasticity of polydomain LCEs under tension is not seen in compression – a result also recently reported by Shaha *et al*.[12] As Warner and Terentjev's famed theoretical semi soft-elastic tensile load curve shape bears a resemblance to that of an ideal dissipator (**Figure 1a**)[13], *i.e.*, a long plateau of constant and finite stress, one would hope to observe and exploit the phenomenon in compression to create highly hysteretic and "ideal" dissipators of impact energy.[14] Interestingly, Shaha *et al*. observed soft-elastic behavior in compression for uniaxial monodomain (macroscopically aligned and anisotropic) LCEs compressed parallel to the molecular symmetry axis, known as "the director." Thus, monodomain LCEs have the potential to act as efficient dissipators of impact energy.

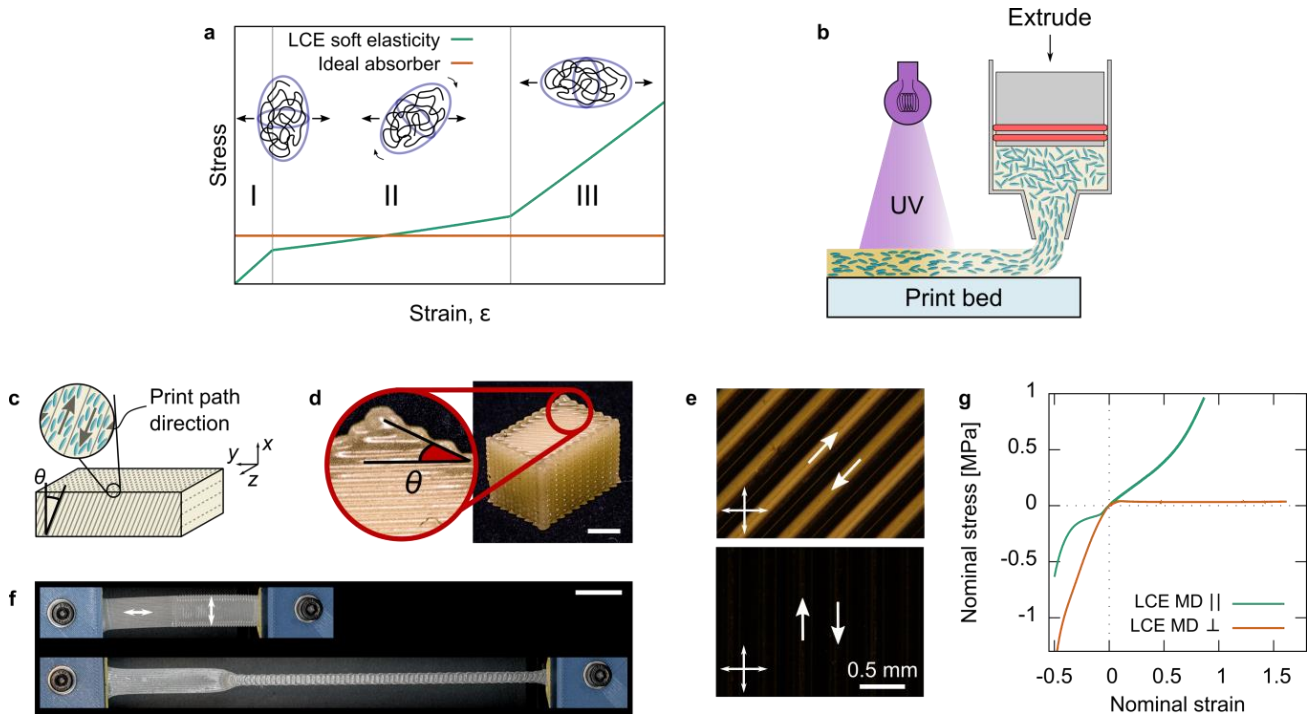

**Figure 1: DIW 3D printing set up and basic material properties. (a)** The rotation of the anisotropic polymer conformation gives rise to the tensile soft-elastic response of LCEs, which bears resemblance to the idealised load curve of a strain-energy absorbing device. **(b)** Liquid-crystal oligomers are shear aligned when extruded through the 3D printer's nozzle and photocrosslinked into an elastomer. **(c)** The direction of print head movement dictates the orientation of the liquid crystal director, thus arbitrarily aligned devices can be constructed. **(d)** An example of a bulk 3D printed LCE, optimisation of the print conditions allows high quality printing of bulk (>cm$^3$) devices. Bar = 5 mm. **(e)** Crossed polarising microscopy of a single printed layer. The uniformity of each image and high contrast between them is indicative of excellent liquid crystalline alignment within printed lines. In the top figure, the striped appearance is caused by cylindrical profile of printed light lensing the transmitted light. Bar = 0.5 mm. **(f)** Mechanical anisotropy and soft-elastic response of the LCE visually demonstrated by straining a bi-strip with domains of parallel and perpendicular orientation. Bar = 10 mm. **(g)** The tensile and compressive mechanical anisotropy of printed LCEs. The soft-elastic response is seen when perpendicular (parallel) oriented samples are stretched (compressed) (n=2).

Historically, fabricating bulk monodomain LCE devices large enough for impact-absorbing applications has been challenging.[15,16] For instance, the two-step LCE synthesis process used by Shaha *et al*. has limited scalability in producing bulk (>cm$^3$) and uniform devices. However, the recent application of direct ink writing (DIW) 3D printing to the fabrication of LCE devices offers a new route to producing arbitrarily sized monodomain LCEs. In DIW, an LC oligomer is shear aligned into a monodomain when extruded through a nozzle – a state which is then fixed by subsequent photo-crosslinking of the oligomers into a network (**Figure 1b**).[15] Line-by-line and layer-by-layer, macroscopic monodomain LCE devices can be fabricated with anisotropy controlled *via* the chosen movements of the print head (**Figure 1c**). Despite the potential of DIW printing to fabricate bulk devices, existing studies have only demonstrated thin shape actuation devices of <10 printed layers or <2 mm thick.[17–23]

In this paper, we DIW print bulk monodomain LCE devices up to 12x12x8 mm$^3$ in size and investigate their anisotropic and dynamic mechanical responses in comparison to equivalent polydomain LCEs and conventional isotropic materials. We show that soft-elasticity in monodomain LCEs offers a fundamentally unique route to enhancing the impact-absorbing behavior of solid elastomeric through quasi-static, high strain rate, and impact testing experiments materials – above and beyond that which conventional isotropic elastomers can provide. Additionally, we show that by programming the print direction within LCE devices, we can introduce and control the nature of buckling responses during compression.

**Results and discussion**

*DIW printing and materials overview*

Large monodomain LCE blocks were DIW-printed after optimizing the printing parameters (**Figure 1d**). The printing remained stable for over the >20 print layers of our devices by selecting the proper ink temperature, extrusion rate, print speed, layer height, and extrusion width (detailed in methods). Fabricating devices on this scale (*i.e.* devices >5 mm thick) is key to our mechanical and impact studies as well as the end-application of LCEs as impact-absorbing materials. The monodomain liquid crystalline alignment of the printed LCE devices is readily seen when viewing a single printed layer (310 µm thick) *via* crossed-polarizing microscopy (**Figure 1e**). When the print direction is oriented at 45°, a highly uniform bright state is seen (**Figure 1e**, top). In the figure, the adjacent printed lines are in contact with one another. The black bands seen are a consequence of the printed lines' cylindrical profile (a result of the circular nozzle outlet) lensing the transmitted light.[15] When the print direction is oriented parallel with one of the polarizers, a dark state is seen (**Figure 1e**, bottom). The appearance of the material and contrast shown in these images indicate a highly uniform state of uniaxial alignment in the printed LCE.

The local mechanical response of DIW-printed LCE devices can be tuned through the design of the printing direction. A bi-stip device (Figure 1f) demonstrates the ability to tune the anisotropic mechanical response of LCEs using DIW printing. Upon straining the device, the perpendicularly oriented region is deformed to a far greater extent than the parallel oriented region – a consequence of soft elasticity.

The magnitude of the DIW-printed LCE's mechanical anisotropy is quantified by the load curve of **Figure 1g**. Tensile strains (upper right quadrant) applied parallel to the print direction (and director) give rise to a conventional elastomeric response. In comparison, for tensile strains up to 1.5 (experimental limit) applied perpendicular to the director, the LCE demonstrates a near-zero modulus across the soft elastic plateau. In compression (lower left quadrant), a similar anisotropic response is seen but with two differences. First, the soft-elastic (classical elastic) response is now seen for strains parallel (perpendicular) to the director. Second, the apparent magnitude of anisotropy is less than that seen in tension, an intuitive result given the maximum nominal strain of -1 in compression. The significant levels of anisotropy and the soft-elastic effect in the bulk DIW-printed LCE devices again indicate high levels of liquid crystal alignment throughout each layer of our printed device.

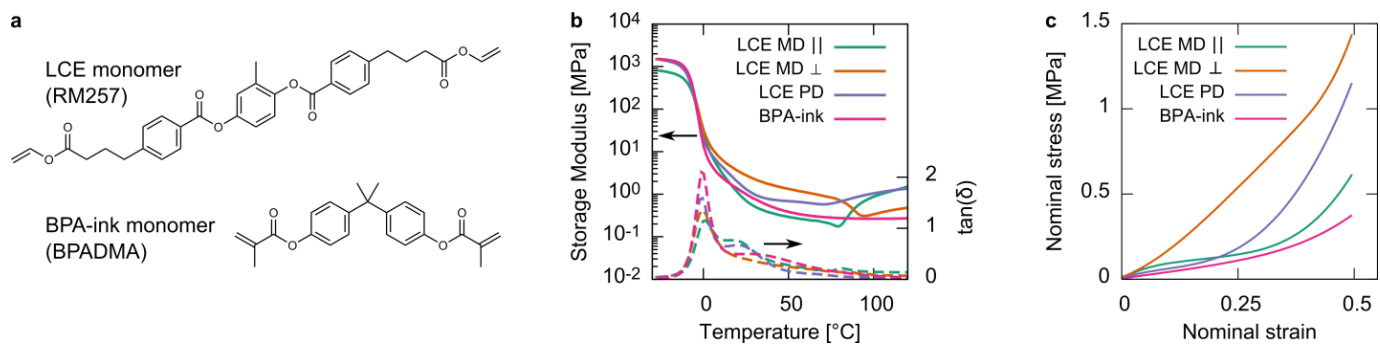

**Figure 2: Key chemical structures and basic mechanical characterisation. (a)** Chemical structures of the diacrylate monomer groups used in this work. RM257 has a stiff core which promotes liquid crystalline ordering. Bisphenol-A dimethacrylate (BPADMA), does not have a rod-like core structure. **(b)** Dynamic mechanical analysis data comparing the storage moduli (solid) and loss ratio, $tan(\delta)$, (dashed) properties of the key materials compared here. (n=2). **(c)** Quasi-static ($10^{-4}$ s$^{-1}$ displacement rate) compressive load curves of all materials compared here. Despite the key structural differences between the LCE MD ∥, LCE PD LCE BPA-ink elastomer, they all have similar low-strain behavior making and can be reasonably compared (n=3).

Here, we compare the dissipative and impact-absorbing capability of our LCE in 3D printed devices when compressed parallel (LCE MD ∥) and perpendicular (LCE MD ⊥) to the director and print direction. We also compare the response of a molded polydomain LCE (LCE PD) of similar chemistry and a DIW-printed isotropic elastomer (BPA elastomer) for further comparison and discussion. The liquid crystallinity of our LCEs results from the use of the diacrylate monomer, RM257, (**Figure 2a**), which is has a central stiff and rod-like core. Our isotropic elastomer material replaces RM257 with bisphenol-A dimethacrylate (BPADMA, **Figure 2a**), which features similar chemical groups as RM257 but does not have a rod-like core.

The LCE PD's and BPA elastomer's crosslink densities were tailored to ensure comparable thermomechanical properties as the DIW printed LCEs. **Figure 2b** shows the storage moduli and $tan(\delta)$ of each material measured using dynamic mechanical analysis (DMA). The peak of each material's $tan(\delta)$ shows that all materials have glass transition temperatures ($T_g$) within 1°C of each other. Additionally, in the rubbery regime above $T_g$, these materials have comparable storage moduli. At room temperature (20°C), the monodomain LCE has an anisotropic storage modulus of 1.1 and 3.6 MPa for small strains parallel and perpendicular to the director respectively. The storage moduli of LCE PD and the BPA elastomer lies between the values for the monodomain LCE at 2.0 and 1.4 MPa respectively. The $tan(\delta)$ curves also show that these materials have elevated $tan(\delta)$ plateaus above $T_g$. While this is expected for LCEs[10], an elevated $tan(\delta)$ was not expected for the BPA elastomer. Overall, the high $tan(\delta)$ of the BPA elastomer makes this an excellent material to compare to the LCEs, allowing us to directly compare the influences of viscoelasticity alone (BPA) versus viscoelasticity coupled with mesogen rotation (LCE).

In addition to the materials already described, we also compare these materials against Sorbothane® Durometer-70 (SD-70, durometer measured on the Shore 00 scale), a commercial polyurethane-based impact absorbing and vibration-isolating material. The compressive load curves in **Figure 2c** show that all materials all have comparable elastomeric behavior at the quasi-static nominal strain rate ($10^{-4}$ s$^{-1}$).

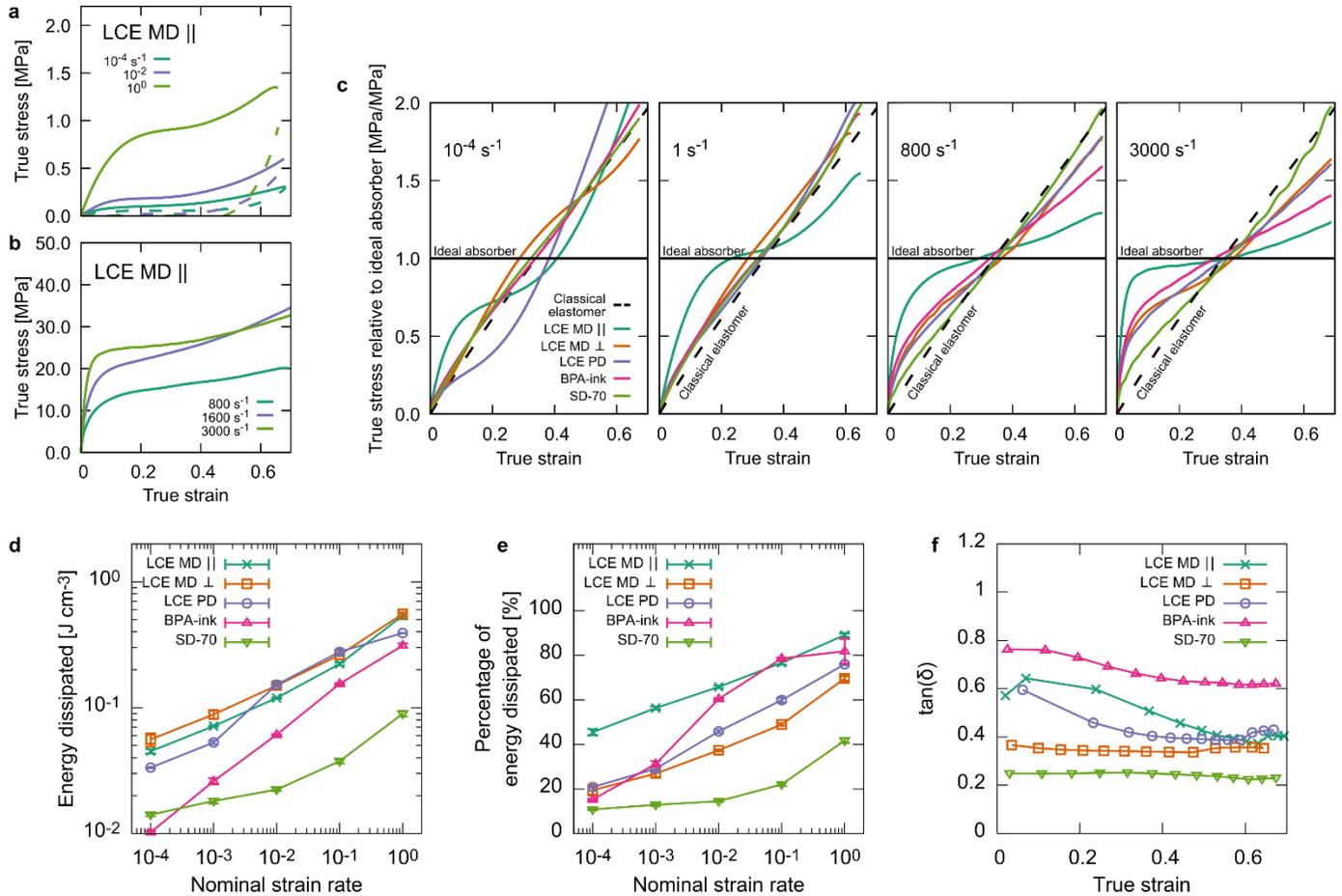

**Figure 3: Strain energy absorption. (a)** Examples of compressive loading (solid) and unloading (dashed) curves for the LCE MD ∥ with nominal strain rates between the quasistatic $10^{-4}$ s$^{-1}$ and the intermediate $10^0$ s$^{-1}$ demonstrating the rate-dependence of the soft-elastic response. The area between the load curves corresponds to dissipated strain energy (n≥3). **(b)** Examples of the compressive load curves at impact-magnitude displacement rates, the rate dependency is still evident with the LCE demonstrating stiffer (softer) elastic behaviours at low (intermediate) strains for increasing displacement rate (n=3). **(c)** Comparisons of each material's compressive behavior at various strain rates relative to the constant-stress behaviour of the ideal dissipator of strain energy. For reference, the expected response of a classical entropic elastomer is also shown. Representative curves are shown from a testing size of n≥3. **(d)** Quantified measurements of the dissipated energy and **(e)** percentage dissipated energy relative to loaded energy for each tested material from quasistatic $10^{-4}$ s$^{-1}$, and the intermediate $10^0$ s$^{-1}$ nominal strain rates. (n≥3) **(f)** The loss ratio, $tan(\delta) = E''/E'$, from small strain (0.1% amplitude) dynamic mechanical analysis tests performed on each sample at different levels of compressive strains (n=5).

*Rate dependency of load curve shape and energy dissipated*

Compressive load curves for LCE MD ∥ over 8 decades of nominal strain rate tested show the material's highly rate-dependent and soft-elastic behavior (**Figures 3a,b**). In each case, samples were loaded to compressive true strains of -0.7 (nominal strains of -0.5, note compressive strains are shown as positive for simplicity). Samples used had a low aspect ratio to avoid any buckling of the samples duing testing

(see methods). True stresses were calculated using a constant volume and ideal deformation (minimal edge effects) assumption by multiplying the nominal stress by the deformation of $\lambda = \epsilon_N + 1$, where $\epsilon_N$ is the nominal strain. We note that these assumptions have their limitations for the presented DIW printed LCEs and BPA elastomers, which have porosities of 18% and 9%, respectively (**Figure S1**). These devices' porosity is a consequence of the cylindrical profile of the extruded inks (a consequence of the circular nozzle outlet) trapping parallel channels of air inside the printed devices. Despite this, these calculations of true stress still provide a realistic insight into the tested materials' mechanical responses, which undergo significant increases in cross-sectional area upon compression. We note that the porous channels do not introduce any anisotropy in the character of the BPA material's dynamic and quasi-static mechanical behavior (**Figure S2b**).

**Figure 3c** shows loading curves for each material at nominal strain rates of $10^{-4}$, 1 ($10^0$), 800 ($8 \times 10^2$), and 3000 ($3 \times 10^3$) $s^{-1}$. In these figures, the true stresses have been normalized against the stress level of an ideal absorber of strain energy (see methods). Presenting the data in this way enables comparison of each material's stress-strain response at each nominal strain rate and how the load curves' characteristic behavior changes with increasing nominal strain rate. For reference, on each graph we also show the expected behavior of a classically elastic and volume conserving material (an incompressible neo-Hookean solid), described by,

$$\sigma_T = \mu(e^{2\epsilon_T} - e^{-\epsilon_T}), (1)$$

where $\mu$ is the usual characteristic rubber modulus, and $\sigma_T$ and $\epsilon_T$ are the true stress and strain, respectively.

From the quasi-static rate of $10^{-4}$ $s^{-1}$ we can identify three types of material behavior. The first type is classical elasticity from materials that did not undergo mesogen rotation (LCE MD ⊥, BPA elastomer, and SD-70). The second is polydomain soft-elasticity displayed by LCE PD. Some soft-elastic effects are visible as the material softens from the classical behavior at a strain of ~0.1, which contrasts our previous work and that of Shaha *et al.* Despite demonstrating a small amount of soft elasticity, the LCE PD displays the worst performance of all materials tested when compared to the ideal absorber. The last type of material behavior is monodomain soft-elasticity with optimised mesogen rotation, shown by LCE MD ∥. In this behavior, the material is initially stiffer than a conventional elastomer and then softens in a plateau – resulting in the closest performance to an ideal absorber.

By the intermediate strain rate of 1 $s^{-1}$, the LCE PD sample no longer displays any soft-elasticity. Therefore, LCE PD behaves, like LCE MD ⊥, BPA, and SD-70, as a classical elastomer – not-optimized for absorbing mechanical energy. By contrast, the rate dependence of the LCE MD ∥ further accentuates the soft-elastic effect and brings the material's performance closer to that of an ideal absorber.

At impact rates of 800 and 3000 $s^{-1}$, LCE MD ⊥, LCE PD and BPA-ink all have similar, slightly improved, responses departing from that of a classical elastomer. However, the LCE MD ∥ still shows a fundamentally different response that is again much closer to ideal behavior. These trends continue to the fastest nominal strain rate tested (3000 $s^{-1}$), where the LCE MD ∥ is converging toward the ideal response, a remarkable behavior for a non-foam material. For additional comparison, we also highlight Wang *et al.*'s similar experiments performed on a polyurea, a material type commonly used in blast protection applications.[24] Their results at impact rates (3300 $s^{-1}$ and greater) show that at low true strains (<0.2), polyureas demonstrate improvements similar to LCE MD ∥ in their load curve shape with increasing nominal strain rate. However, at true strains greater than ~0.2, the polyurea showed significant stiffening and appeared to "bottom out" at relatively low strains. Therefore, monodomain soft-elastic LCEs show enhanced characteristics for energy absorption high-rate impacts compared to common incumbent materials.

In addition to the loading and storage of energy during a compressive deformation, for impact-absorbing applications it is also important to consider the magnitude (**Figure 3d**) and percentage (**Figure 3e**) of loaded energy that is dissipated. For example, a highly hysteretic and dissipative material is not necessarily a more optimal impact-absorber when compared to an elastic material if it is significantly softer and therefore does not dissipate a large enough magnitude of energy for a given application.

**Figure 3d** shows that from rates of $10^{-4}$ to $1\ s^{-1}$, the LCE samples dissipate similar magnitudes of strain energy and have similar rate dependencies. Even though the LCE MD ∥ has the softest compressive response (**Figure 2c**), it dissipates a comparable magnitude of energy as the stiffer LCE samples due to its highly soft elastic and hysteretic behavior (**Figure 3a**). Furthermore, the LCE MD ∥ demonstrates this similar performance while minimizing peak stresses and acting the closest to an ideal absorber across strain rates (**Figure 3c**). In comparison to the LCEs, the BPA elastomer dissipates a significantly lower magnitude of energy at the quasistatic rate of $10^{-4}\ s^{-1}$ (~25% of the LCE's dissipatation), however this increases with a greater strain rate dependency (~50% of that LCE MD ∥ and LCE MD ⊥ dissipate at $1\ s^{-1}$). At $10^{-4}\ s^{-1}$, SD-70 also dissipates significantly less energy than the LCEs, but unlike the BPA-elastomer this disparity increases with strain rate.

**Figure 3e** demonstrates how LCE MD ∥ dissipates the highest percentage of strain energy. Across strain rates, and particularly at low rates, LCE MD ∥ dissipates a far greater proportion of the loaded energy, from 45±1% at a rate of $10^{-4}\ s^{-1}$, rising to 89.1±0.8% at $1\ s^{-1}$. By comparison, LCE MD ⊥ and LCE PD both dissipate ~20% and ~73% of loaded energy at rates of $10^{-4}$ and $1\ s^{-1}$, respectively. While at room temperature, these slower rates do not relate to impact conditions, they provide an insight to the enhanced dissipative behavior at higher temperatures *via* the time-temperature superposition principle. The BPA shows a curiously different response. Looking back to **figure 2c**, we see that at quasi-static nominal strain rates, the BPA elastomer has a similar stress-strain loading curve as LCE MD ∥ (i.e. it loads a similar magnitude of energy), however, the BPA elastomer instead behaves quite elastically and dissipates only 15.2±0.3% of the loaded energy. This percentage is similar to LCE MD ⊥, and LCE PD – which offer little or limited mesogen rotation capability, but is 67% less than LCE MD ∥ – which has a large capacity for mesogen rotation (**figure 3d**). However, by $1\ s^{-1}$, the BPA elastomer can dissipate a similar percentage of loaded energy (82±6%) to LCE MD ∥ (89.1±0.8%). **Figure 3e** also shows that SD-70 demonstrates by far the lowest dissipative capacity of all materials tested.

From the load curves, we can deduce that at low strains, LCE MD ∥ can accumulate more strain energy (faster than the material can relax) than any other material tested. The stress level achieved by LCE MD ∥ is then somewhat maintained during the soft-elastic plateau, where there is a greater balance in the rate at which strain energy builds and dissipates. The fast reduction in stress as LCE MD ∥ unloads means LCE MD ∥ had already dissipated much of its stored energy at the point at which unloading begins. While this implies viscous and plastic behavior, we note that the LCE MD ∥ sample returned to its original shape within a few minutes of each test's completion, and the responses were almost identical upon repeated tests. The ability to dissipate loaded energy quickly and effectively is essential for impact-absorbing applications where any energy not dissipated can be returned as kinetic energy loading to rebounding.

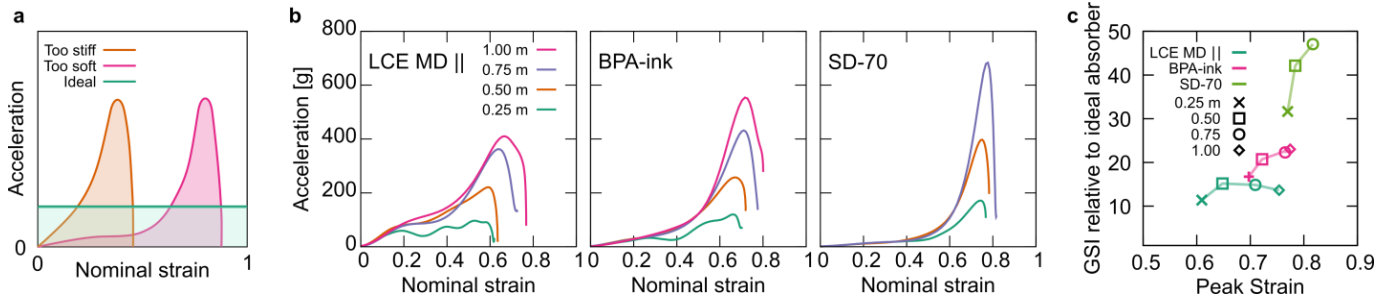

**Figure 4 Impact performance a** Illustration of drop test responses for typical non-porous materials of different stiffnesses, and that of the idealised impact absorber which would fully compress at a constant acceleration. **b** Representative readings from an accelerometer mounted on a 2 kg mass as it is dropped from various heights onto devices of LCE MD ∥, BPA elastomer and SD-70. **c** A summary of drop test results in an Ashby-style plot. Plotting, for each material and drop height, the peak strain against the Gadd severity index (equation 3) reveals each material's balance of minimising the severity of impacts and their capability to increase their maximum strain in response to impacts of greater intensity. n=3 for all data shown.

The BPA elastomer's comparatively sharp increase in percentage dissipated energy at a threshold rate of ~$10^{-2}$ $s^{-1}$ indicates a change in the thermomechanical response of the BPA elastomer in a way which appears to make it suitable for dissipating mechanical energy. This is attributed the time-temperature superposition principle, in which an increase in strain rate is shifting the material closer to its glass transition, from an elastomeric to a leathery response with increased viscous effects. However, this also means that at increasing temperatures, one would expect the BPA elastomer's dissipative capability to notably drop off when compared to the LCEs as the material quickly increases in its elasticity.

By comparing the energy dissipated as a function of strain rate (**figure 3e**), we can start to quantify the contributions of mesogen rotation and liquid-crystallinity in elastomers. The LCE materials all have equivalent chemical compositions but different director orientations to the axis of loading. The LCE MD ∥ samples demonstrated the highest percentage of energy dissipated, illustrating the benefit of LCEs aligned for optimized mesogen rotation. When comparing the LCE MD ∥ to non-mesogenic elastomers, the contributions of mesogen rotation can be clearly seen at low strain rates where viscoelastic effects are minimized. At higher strain rates, the non-mesogenic BPA and SD-70 polymers demonstrate increased viscous effects due to time-temperature superposition and increase energy dissipation. Overall, the tailored structure of the LCE MD ∥ sample provides a superior combination of optimized mesogen rotation and viscoelasticity to demonstrate enhanced energy dissipation across a wide range of test conditions.

This inferred behavior is supported by **Figure 3f,** which shows the *tan(δ)* of each material at various levels of compressive strain. The 1Hz sinusoidal strains of 0.1% amplitude correspond to a root mean square nominal strain rate of $5 \times 10^{-3}$ $s^{-1}$. The *tan(δ)* of all the materials tested have differing extents of strain-dependency. For the LCEs, the differences can be linked to the presence of soft-elastic effects. For the non-soft elastic LCE MD ⊥, *tan(δ)* remains constant with strain at a base level of ~0.35. For the polydomain soft-elastic LCE PD, *tan(δ)* is initially elevated at ~0.6, but then quickly converges to the base level. The monodomain soft-elastic LCE MD ∥ starts with a *tan(δ)* similar to that of LCE PD, however this is maintained for a greater range of strains before converging to the based level. Given that the monodomain soft-elasticity in LCE MD ∥ allows for greater extent of mesogen rotation with strain than the polydomain soft-elasticity in LCE PD, we conclude that mesogen rotation has a significant impact on dissipation in LCEs and enhances the dissipative performance over non-soft elastic (LCE MD ⊥) and conventional (SD-70) elastomers. A notable surprise shown by **figure 3f** is that the BPA elastomer has an exceptionally high *tan*(δ), between ~0.8 at low strains and ~0.6 at high strains, greater than for any other material tested here. Despite having the highest tan(δ) values throughout compression, the LCE MD ∥ still outperformed the BPA elastomer in terms of magnitude and percentage of energy dissipated, highlighting the importance of optimised mesogen rotation compared to traditional viscoelasticity alone.

*Impact performance*

Rate-dependent compressive mechanical testing has shown that monodomain soft elasticity offers several enhancements in an elastomer's ability to dissipate energy. Next, we consider the performance during impacts simulated by drop testing (**Figure 4**). Impact testing is fundamentally different from compressive testing as the strain rate of the sample is not constant throughout the duration of the test. The LCE MD ∥, BPA elastomer, and SD-70 materials are compared to explore how the promising characteristics of monodomain soft elasticity translate to impact behavior. LCE MD ⊥ and PD samples are not presented due to their structure not taking advantage of soft elasticity under compression.

For illustrative purposes, **Figure 4a** shows the impact response for conventional soft and stiff elastomers compared to ideal behavior. The limitation of conventional elastomers is that they will often slow the impacting object *via* a sharp peak in acceleration over a narrow strain range. Conversely, the theoretical ideal shock absorber instead provides a constant acceleration over the entire possible strain range, fully compressing to 100% – thus minimizing the peak acceleration experienced. For simplicity, we do not consider the consequences of "jerk" – the rate of change of acceleration with respect to time, except to note that step-like changes in acceleration are not always desirable during impacts.

**Figure 4b** shows the acceleration experienced by the dropped mass as it impacts and strains the samples. A list of drop heights, impact speeds, energy densities, and initial impact nominal strain rates in these tests is shown in **Table S1**. The LCE MD ∥ demonstrates plateaus in the accelerations at intermediate strains that we can reasonably conclude are manifestations of monodomain soft elasticity in impact conditions. These soft elastic plateaus enable a more uniform deceleration of the dropped mass, giving a response closer to that of the ideal absorber shown in **Figure 4a**. Additionally, the acceleration level for the soft-elastic plateau increases with impact energy and acts as a mechanism to offset increases in the peak acceleration. While all materials tested here are clearly viscoelastic (**Figure 3**), only LCE MD ∥ translates these rate-dependent effects to their mechanical response in drop tests. This is a unique, passively adaptive mechanism whereby the material behaves stiffer in response to impacts of increasing intensity. By comparison, the BPA elastomer and SD-70 lack any intermediate plateau in accelerations and instead demonstrate a 'too-soft' response as the deceleration is confined to a sharp peak in acceleration at strains of ~0.7 and which undergoes greater increases in peak height compared to the curves for LCE MD ∥.

For head-impacts, the severity of an impact and the probability of a person suffering a concussion significantly increases nonlinearly with peak acceleration.[25] These effects can be encapsulated and quantified using the Gadd Severity Index (GSI) which is calculated *via* the following integral[26]

$$\int_0^t dt\, a(t)^{2.5}, \quad (3)$$

where $a(t)$ is the acceleration of the impact as a function of time. The 2.5 powers in the GSI penalizes sharp peaks in, and high values of acceleration. Here, we perform the integral for times from the start of the impact until the velocity of the dropped mass is zero. Each GSI was normalized to the GSI of an idealized absorber (see methods).

**Figure 4c** plots, for each test condition, the GSI against the peak strain. In response to the lowest intensity impacts, the LCE MD ∥ undergoes a significantly lower strain (0.6) than the BPA elastomer and SD-70 (0.7 and 0.8 respectively) while also experiencing the lowest GSI – 30% lower than that of the BPA elastomer. As the drop height increased, the LCE MD ∥ samples maintained a relatively constant normalized GSI performance, remaining below 16. However, the maximum strain significantly increases by 0.15 – the most of any of the materials tested. This increase in maximum strain, along with the increased plateau height, offsets the increase in the peak acceleration and is responsible for the relatively constant GSI performance. At the maximum impact intensity tested, the GSI for the monodomain soft elastic LCE MD ∥ performs 40% better than the BPA elastomer. The lesser performance of the BPA elastomer is quite curious given the DMA data of **Figure 3f** which would suggest the material to be an effective absorber of mechanical energy. The resolution of this apparent inconsistency is that while at intermediate-to-high speed deformations the BPA material can effectively dissipate the energy which is loaded, it is fundamentally limited by being too soft early in it deformation and so it fundamentally has limited characteristics suited for safely absorbing impact energy. Lastly, the commercial impact-absorbing material, SD-70, is much too soft to absorb impact energy of this magnitude as with increasing drop height, the maximum strain barely increases while the peak acceleration, and hence GSI increases significantly – this is characteristic of the material "bottoming-out". Note that for SD-70, drop tests from 1.00 m were not performed to protect the experimental apparatus.

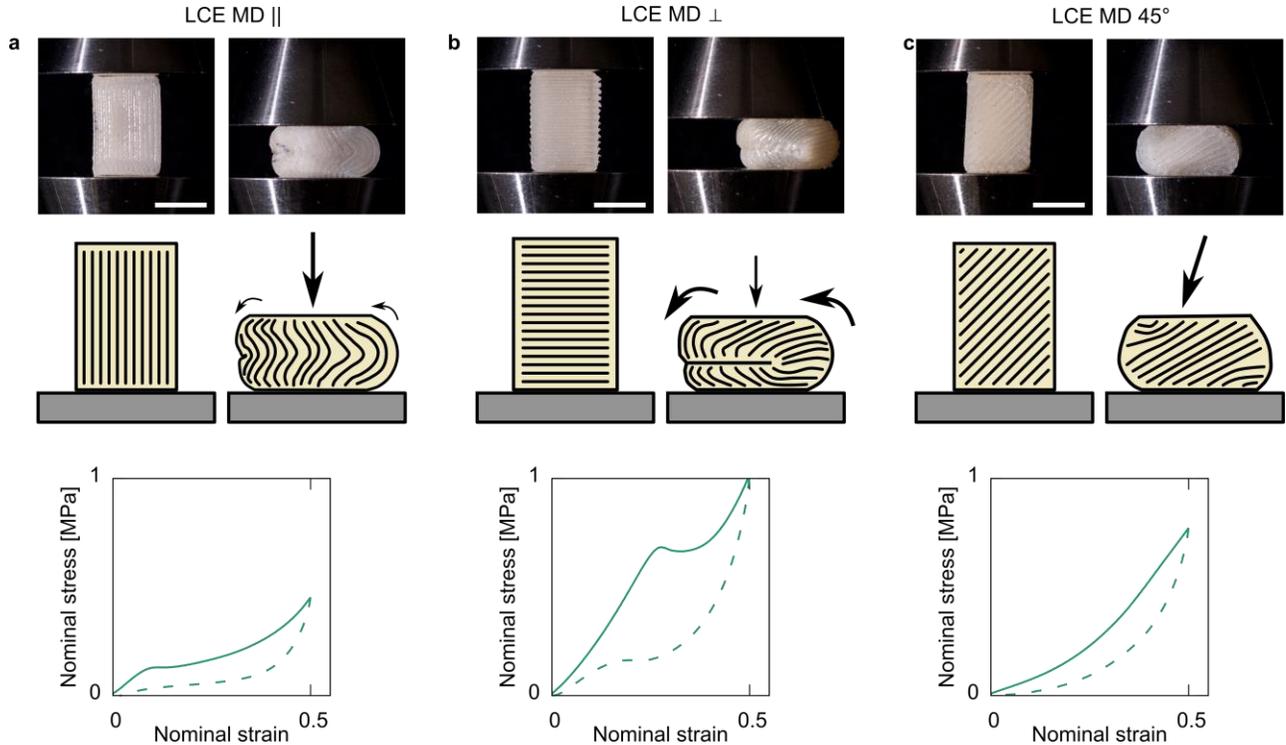

**Figure 5 Anisotropy-controlled buckling** By compressing high aspect ratio (height > in cross sectional dimension) monodomain LCE devices with the director oriented (**a**) parallel, (**b**) perpendicular and (**c**) at 45° to the compression axis, we can further control the buckling response and load curve shape of the DIW printed LCE devices. The figures show photographs to devices in their strained and unstrained states along will illustrations of the director profile in each state – traced from the print lines visible in each device, and the compressive nominal stress – nominal strain load curves recorded for each test (n=1 for each test). Bars are 10 mm.

*Controlling buckling deformations with anisotropy*

The richness in the mechanical behaviors that can be achieved with anisotropic monodomain LCEs evidently show great promise in impact-mitigating technologies. Until this point we have solely compared the inherent material characteristics in low aspect ratio devices where buckling is avoided. However, buckling deformations offer additional modes for the dissipation of mechanical energy.[27] We now demonstrate the additional levels of control over compressive mechanical behavior possible in DIW-printed monodomain LCE devices in high-aspect ratio devices where buckling effects are likely (see methods).The DIW-printing process imparts liquid-crystal alignment into the material to control anisotropy and buckling behavior (**Figure 5**). Three high aspect-ratio LCE pillars were tested with the print pathways 0, 45, and 90° to the direction of compression. The photographs show the devices at 0 and 50% nominal strains, and the illustrations trace the director profile *via* the visible printed lines. Compressing along the director (**Figure 5a**) yields the LCE MD ∥ response seen throughout our results, where soft elasticity is analogous to microscopic buckling within the device. Compressing perpendicular to the director (**Figure 5b**) initially gives the LCE MD ⊥ response seen in figure 3. However, past a nominal strain of 0.25, the high aspect-ratio sample undergoes a macroscopic buckling instability, caused by a low shear modulus from the shearing of printed layers along the director. Lastly, for compressions applied at 45° to the director, all buckling effects are suppressed, and the devices demonstrates a more classical elastic response despite its high aspect-ratio geometry (**Figure 5c**). This additional degree of buckling complexity that one can introduce and control *via* the print orientations opens the door to mimicking the mechanics of biological materials such as the heterogeneous and anisotropic intervertebral disc.[28]

*Discussion*

In this work, we explored the effects of soft elasticity in bulk monodomain LCEs under compression, which has potential use in impact absorbing devices. Through careful optimization of DIW print conditions, we were able to fabricate (to the best of our knowledge) the largest monodomain LCE devices reported to date, and which allowed us to conduct an exceptionally in-depth study of LCE compressive mechanical properties. By directing compressions along the liquid crystalline director, LCE dissipative devices are capable of dissipating large quantities of strain energy at relatively constant levels of stress – a close-to-ideal behavior which avoids peaks in stress, and almost all of which (90% at a strain rate of 1 $s^{-1}$) is quickly dissipated. In practically all metrics tested, the LCE MD ∥ samples outperformed a chemically identical LCE loaded perpendicular to the director, a thermodynamically equivalent polydomain LCE, and an isotropic BPA-ink elastomer with a similar glass transition temperature. We attribute the performance of the LCE MD ∥ to its most significant material differences to the other materials – its monodomain and soft elastic nature.

The DMA results for our comparator conventional BPA elastomer succinctly demonstrates how the performance of a materials as dissipators of large-strain or impact energy is not just determined by a the material exhibiting a high $tan(\delta)$ throughout its load curve. While this clearly can be highly optimized (the BPA elastomer had the greatest $tan(\delta)$ of all materials tested), steps must also be taken to ensure the material loads energy in an optimized manner. Monodomain soft elasticity is evidently an effective mechanism to realize improved impact-absorbing behavior in solid elastomers with the additional rate-dependency offering a material that performs consistently over impacts of differing intensities.

There is much to be explored in future studies. First, we note the LCEs studied here are synthesized from liquid crystalline monomers that were optimized for the displays industry and chosen here (along with the other components) for their low-cost and availability. Future studies should seek to optimize material design and understand how LCE dynamics, *i.e*. the different relaxation processes and their timescales, affect load curve shape, rate dependency and dissipation. This research will undoubtedly lead to LCE devices of even greater performances than those reported here. At the same time, there is evidently much richness yet to be explored in understanding and exploiting the anisotropy-controlled buckling deformations of LCEs. One can easily envisage how such deformations could be used to further optimize the nature of LCE deformations for a wide range of impact scenarios, for instance, in controlling rotational accelerations associated with oblique impacts.

In short, we have shown here that coupling the long-famed anisotropic non-linearity of LCEs with DIW 3D printing opens enormous application potential of impact absorbing LCE devices, with much physics and mechanics yet to be explored.

**Conclusion**

Here, we have demonstrated optimized DIW 3D printing of LCEs which we used to create the largest known monodomain devices to date. By testing these in compression over eight decades of strain rate – spanning quasi-static and impact rates, we have shown how the monodomain soft elastic mechanism enhances an elastomer's capability to load and dissipate mechanical impact energy. The monodomain soft elastic LCEs demonstrated the closest stress-strain response to that of an ideal absorber and its rate dependency led to a relatively consistent performance over drop tests of differing intensities. We also showed how compressive deformations can be further enhanced by using the DIW printing direction to control the buckling characteristics of the LCE. This work brings the realization of LCE devices closer as it demonstrates a scalable and industrially-realistic fabrication method along with enhanced material behavior not seen in conventional elastomers.

**Methods**

*Materials and oligomer synthesis*

Acrylate-capped LC oligomers were synthesized using 4-(3-acryloyloxypropyloxy)benzoic acid 2-methyl-1,4-phenylene ester (RM257, CAS 174063-87-7), 2,2'-(ethylenedioxy)diethanethiol (EDDT, CAS 4970-87-7), butylated hydroxytoluene (BHT, CAS 128-37-0), 2-Hydroxy-4'-(2-hydroxyethoxy)-2-methylpropiophenone (HHMP, CAS 106797-53-9) and N,N,N',N'',N''-Pentamethyldiethylenetriamine (PMDETA, CAS 3030-47-5). Polydomain LCEs were synthesized with the same chemicals, but with the addition of pentaerythritol tetrakis(3-mercaptopropionate) (PETMP, CAS 7575-23-7). The non-mesogenic acrylate capped oligomers were synthesized with the same components as the LC oligomer, but with RM257 replaced with the non-mesogenic diacrylate bisphenol A dimethacrylate (BPADMA, CAS 3253-39-2). RM257 was purchased from Wilshire Technologies, all other components were purchased from Sigma Aldrich and all components were used as received. Chemical structures of RM257 and BPADMA are shown in all components are shown in figure S3. Sorbothane Duro 70 was purchased from Isolate it! And cut to size for testing.

Liquid crystalline oligomers were synthesized *via* a base-catalyzed thiol-Michael click-reaction, described in detail elsewhere.[15] Briefly, BHT (radical-inhibitor, 1.4 mol. % of total reactants), RM257 (diacrylate mesogenic monomer, 51.1 mol. %) and HHMP (UV-radical photoinitiator, 1.9 mol. %) were added to a glass vial and melted together in a water bath set at 70°C. The melted components were thoroughly mixed, and bubbles/dissolved gases removed via vacuum before EDDT (dithiol spacer monomer, 44.4 mol. %) and PMDETA (base catalyst, 1.2 mol. %) were added and the mixture again mixed and degassed. The mixture was then transferred to the DIW-printing barrels and left in an oven set at 70°C for half an hour to start the Michael addition (figure S3). The barrel was then left at ambient temperature and protected from light for 2 days before printing. The chosen ratio of RM257:EDDT (1.000:0.870) ensured oligomers were acrylate capped and therefore would undergo crosslinking during 3D printing.

The LCE PD was synthesized *via* similar method and composition, with the addition of the tetra-functional thiol crosslinker. We used a material of RM257:EDDT:PETMP = 1.000:0.871:0.065 as this had an even balance of acrylate groups to thiol groups – ensuring that after the Michael addition (assuming complete conversion) no excess acrylate or thiol groups remained. In this synthesis, PETMP was added at the same time as the EDDT and once all components were combined together, the mixture was poured into molds as opposed to the printing barrels.

For the synthesis of the non-liquid crystalline BPA-ink, the same oligomerization process used for the LC oligomer was used except for the following differences. First the non-mesogenic diacrylate monomer BPADMA was used in place of RM257 and in a ratio of BPADMA:EDDT = 1:0.96. Second, the quantities used of BHT, HHMP and PMDETA were 4.0, 4.0 and 4.1 mol. % of total reactants respectively.

*Direct-ink writing 3D printing*

DIW printing was performed using a Hyrel Engine HR 3D printer equipped with a KRA-2 print head for heating and extruding LC oligomers along directed print paths. Barrels containing printable oligomer were installed in the KRA print head which was set at 65°C for the LC-ink (left for an hour prior to printing for equilibration) and kept at ambient conditions for the BPA-ink (due to is significantly lower viscosity). During printing, materials were extruded through a Tecdia Arque-S 5060 nozzle which had an internal diameter of 500 μm at the nozzle tip. G-code toolpaths controlling the print head's motion, printer settings and volumetric rate of material extrusion were created using in-house developed *python* scripts which also aided tuning of the print parameters (process described in the ESI). During extrusion, the extruded material was exposed to UV light from LEDs surrounding the nozzle. Post-printing, the devices (typically 12x8x8 mm) were fully cured through exposure to high intensity UV light in a UVP CL-1000 (Ultraviolet Crosslinkers, Upland, CA, USA) chamber for 2 hours. Devices used in tensile stress-strain and DMA tests had far greater surface area-to-volume ratios and so were post-cured for 30 minutes. All printed devices were periodically rotated during post-curing to ensure even exposure.

Print conditions for our LCE were optimized through printing a series of matrices tuning, in turn, the various print parameters. First, we printed meanders of single lines - simultaneously optimizing for the volumetric extrusion rate and the nozzle height above the print surface. Liquid crystalline alignment quality was assessed via polarizing microscopy, with the parameters offering the greatest apparent uniformity in, and contrast between the bright and dark states chosen as the ideal parameters. Next, using these parameters we printed series of meandering lines of different spacing between print lines until the print lines were close enough to bond to each other form a single printed sheet. Care was taken not to print lines too close to one another – which would diminish the level of liquid crystalline alignment present. By measuring the thickness of the printed sheets, we deduced the ideal layer height to use for multi-layered devices.

*Dynamic Mechanical Analysis*

Striped of monodomain LCE and BPA-ink elastomers were cut from larger printed sheets of printed layers, with the long edge either parallel or perpendicular to the print orientation as necessary. Strips had dimensions of ~25x5x1 mm with the gauge length being ~ 15 mm once clamped. Additionally, strips of polydomain LCE were cut from molded sheets of ~1 mm thickness.

Iso-frequency DMA temperature sweeps of the LCEs and BPA-ink elastomer was performed using a TA Instruments DMA Q800 equipped with an ACS-2 refrigerated air supply. Samples were loaded with a 0.01 N preload force (force tracking at 120% enabled) and were subject to 1Hz oscillations of 0.1% strain amplitude. Samples were heated to 130°C and allowed to equilibrate for 10 minutes to erase their thermal history before data were collected during a temperature sweep to -30°C at 2°C min$^{-1}$.

*Slow/intermediate rate mechanical and $tan(\delta)$ testing*

Compressive mechanical tests were performed using a TA Instruments Electroforce 3230 equipped with a ±450 N load cell. Printed LCE and BPA-ink, LCE PD and SD-70 devices of dimensions ~12x8 mm$^2$ cross-sectional area and 8 mm height were prepared and in the case of printed devices, lightly sanded after freezing to give flat surfaces. The sample dimensions following sanding were used in analysis.

For comparisons of each material's compressive load curve and rate dependency, each device was tested by application of an initial 10 kPa preload stress (~1 N and loaded at 0.01 N s-1) to ensure the platen was in contact with the top of the sample. After dwelling at this position for 600 s, the samples were loaded to 50% nominal strain and unloaded 0% to strain at strain rates of 10$^{-4}$ (taken as quasi-static), 10$^{-3}$, 10$^{-2}$, 0.1 and 1 s$^{-1}$. From the data collected, the dissipated strain energy density (area between loading and unloading curves) and percentage dissipated energy (area between load curves relative to area under loading curve) were calculated. For display in figure 3, the load curves of repeated tests were averaged.

Measurements of the loss tangent, $tan(\delta)$, at different strains across each material's compressive load curve were performed using the Electroforce's DMA Application. Samples were loaded with a series of incremental forces (chosen based on each material's stiffness to 50% compressive strain). At each load increment, the samples were allowed to stress relax for 10 minutes before being subjected to 0.1% strain amplitude oscillations for DMA tests. Using each incremental load, the sample strain was extracted using the quasi-static compressive load curves measured for each material.

The buckling characteristics of higher aspect ratio (8x8 mm$^2$, cross-sectional area and 12 mm height) printed devices was assessed by subjecting the samples to a 10 kPa preload force (loaded at 0.01 N s$^{-1}$) and then loaded to 50% and unloaded at a rate of 10$^{-3}$ s$^{-1}$. Photographs were taken of each sample in their unstrained and maximally strained states and from which the director orientation (on the face seen by the camera) could be traced using the print lines which were visible.

*Kolsky bar testing*

A Kolsky (also called split-Hopkinson) bar was used to measure the high nominal strain rate compression response of the materials beyond 1 s$^{-1}$. The Kolsky compression bar setup is composed of three axially aligned rods, the striker, incident, and transmission bars. The specimen is sandwiched between the incident and transmission bars in a stress-free state prior to the start of the experiment. The Kolsky bar is actuated when the striker bar is accelerated using compressed gas from a gun barrel. When the striker bar impacts the end of the incident bar, an incident stress wave is generated that propagates along the bar until it reaches the sample. The specimen is compressed at a high deformation rate when the incident wave arrives at the end of the incident bar. Part of the wave is transmitted through the sample into the transmission bar, and part is reflected in the incident bar as a reflected wave. Strain gages mounted on the incident and transmission bars allow measurement of the specimen nominal stress, strain, and strain rate according to.[29] Pulse shaping is a critical step in achieving a constant nominal strain rate deformation and an equilibrated stress in the sample, both of which are required for a valid experiment. Small disks of annealed copper, or "pulse shapers" are placed on the impact end of the incident bar. The dimensions of the disks are designed to vary the profile of the incident pulse to achieve constant strain rate in the sample.

The dimensions of the Kolsky compression samples were approximately 2 x 5 x 5 mm$^3$ (2 mm thick), dimensions which aided the stress equilibration process. Experiments were carried out at nominal strain rates of 800, 1600, and 3000 s$^{-1}$ with n=3 at each condition. Samples were subjected to a minimum of 50% nominal strain at each condition for comparison with quasi-static experiments. For display in figure 3, the load curves of repeated tests were averaged.

*True strain relative to ideal absorber*

First, for each material and at each nominal strain rate (using their averaged load curves), the amount of loaded strain energy was calculated by integrating the load curves between true strains of 0 and 0.7. Diving this by a true strain of 0.7 gives the idealized true stress level of an ideal absorber - which would absorb the same quantity of strain energy but at constant true stress. The normalized load curves were then generated by dividing the stresses from the averaged load curves by their respective idealized true stress level.

*Drop testing*

Drop tests were performed using an in-house built test frame. A linear accelerometer, attached to a 2 kg mass mounted on a guiding linear rail, was dropped on samples of typical dimensions 12x12 mm$^2$ cross-sectional area and 8 mm height. Each material was subjected to drop tests with the mass dropped from heights of 0.25, 0.50, 0.75 and 1.00 m (due to its softness, SD-70 was not subjected to the drop from 1.00 m). Table S1 shows, for each drop height, the impact speed, initial nominal strain rate and impact energy density. Accelerometer data was passed through a SAE 1000 filter and began recording data once a change in acceleration of +1 g (=+9.812 m s$^{-2}$) was detected – a displacement position taken as the top of the impacted sample. The impact velocity was deduced from a separately determined calibration curve linking drop height to impact velocity. Continued displacement was calculated using the impact velocity and twice integration of the accelerometer's readings.

GSI for ideal absorbers, capable of constant deceleration to 100% compressive nominal strain, were calculated using the impact velocity and height of each sample to determine the ideal impact time and deceleration value. The GSI of each sample at each drop height was compared to this value of the idealized absorber.

**Acknowledgements**

DM thanks support from the English Speaking Union through a Lindemann Fellowship and the Leverhulme Trust through an Early Career Fellowship. This work was supported in part through the Laboratory Directed Research and Development program at Sandia National Laboratories. Sandia National Laboratories is a multimission laboratory managed and operated by National Technology & Engineering Solutions of Sandia, LLC, a wholly owned subsidiary of Honeywell International Inc., for the U.S. Department of Energy's National Nuclear Security Administration under contract DE-NA0003525. This paper describes objective technical results and analysis. Any subjective views or opinions that might be expressed in the paper do not necessarily represent the views of the U.S. Department of Energy or the United States Government. SAND2021-3472 O.

**Conflict of interest**

C.M.Y. and D.M. have potential conflict of interest since they own equity in companies that are trying to commercialize LCE products.

# Supplementary Information for Optimizing dissipation through the soft-elasticity of 3D printed liquid crystal elastomers.

D. Mistry[1,2]*, N. Traugutt[1], B. Sanborn[3], R. Volpe[1], Lillian Chatham[4], Risheng Zhou[1], B Song[3], Kai Yu[1], K. Long[3], C. Yakacki[1]*

[1]Department of Mechanical Engineering, University of Colorado, Denver, Denver, CO, 80204, USA.
[2]Present address: School of Physics and Astronomy, University of Leeds, Leeds, LS2 9JT, UK.
[2]Materials and Failure Modeling Department, Sandia National Laboratories, Albuquerque, NM, 87123, USA
[4]Impressio Inc., 12635 E. Montview Blvd, Suite 214, Aurora, CO 80045, USA

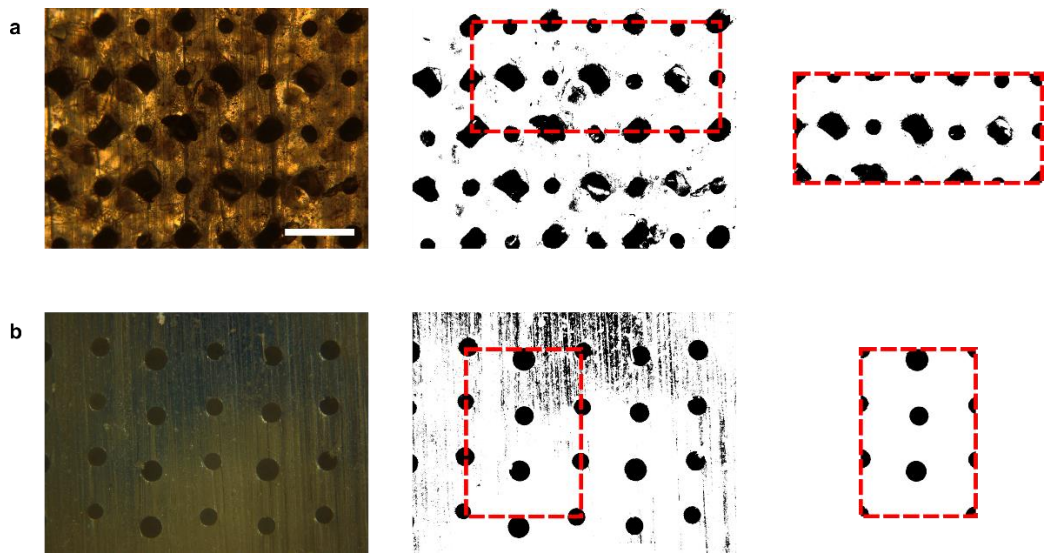

**Supplementary figure 1 Porosity of DIW-printed devices** The DIW process trapped channels of air, parallel to the print direction inside printed devices. Thin example slices of the plane perpendicular to the print direction for the printed LCEs (**a**) and BPA-ink elastomer (**b**) are shown, viewed via reflection microscopy. The black regions are slices of the porous channels. By processing these images to separate air and elastomer regions, and defining a representative unit cell, approximation of the devices' porosity could be made. The LCE (BPA-ink elastomer) was calculated to have a porosity of 18 % (9 %). Bar is 1 mm.

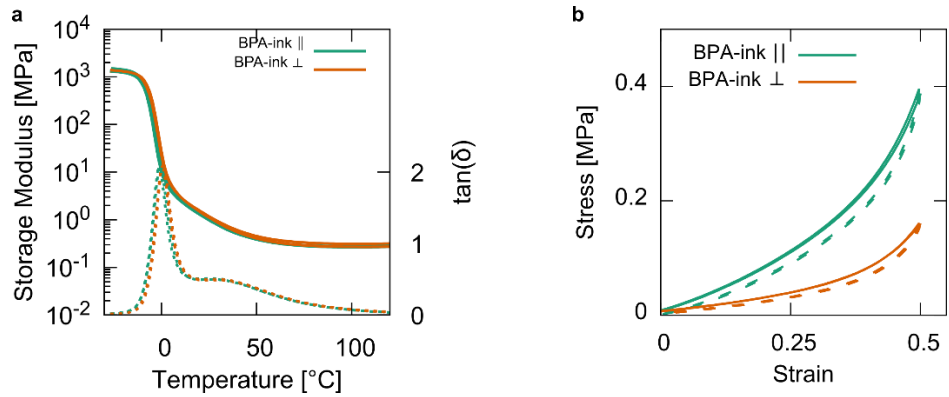

**Supplementary figure 2 Isotropy of BPA-ink elastomer. a** DMA temperature-sweeps and, **b**, quasi-static compression tests ($10^{-4}$ s$^{-1}$) of the 3D printed BPA-ink elastomer devices. In both cases, samples were tested for strains applied parallel and perpendicular to the printed orientation. In **a,** the material's behavior was identical. In **b** the parallel sample behaved stiffer, however the character of the load curves is the same. The figures confirm the fundamental isotropy of the BPA-ink elastomer with respect to printed orientation.

EDDT 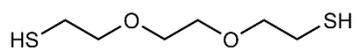

PETMP 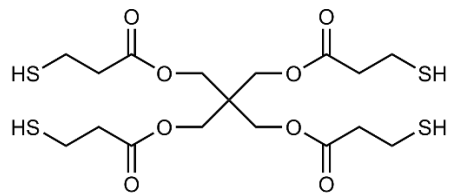

PMDETA 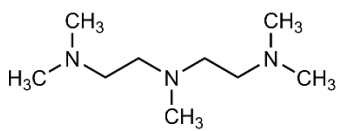

HHMP 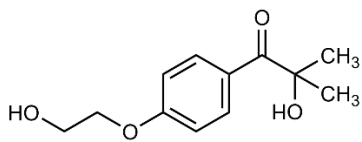

BHT 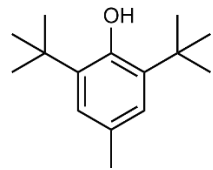

**Supplementary Figure 3** Structures of chemicals used in this work. Additional structures of the liquid crystalline and non-liquid crystalline diacrylate monomers also used are shown in figure 2a.

**Table S1 Drop heights and equivalent parameters of 2 kg drop tests.**

| Drop height, m | Impact speed, m s$^{-1}$ | Initial nominal strain rate, s$^{-1}$ | Impact energy density, J cm$^{-3}$ |
|---|---|---|---|
| 0.25 | 2.1 | 270 | 4.0 |
| 0.50 | 2.9 | 380 | 8.0 |
| 0.75 | 3.6 | 470 | 12.0 |
| 1.00 | 4.2 | 540 | 16.0 |